\documentclass[12pt]{iopart}
\usepackage{epsfig}
\usepackage{citesort}

\newcommand{\gtrsim}{\,\rlap{\lower3.7pt\hbox{$\mathchar\sim$}}
\raise1pt\hbox{$>$}\,}
\newcommand{\lesssim}{\,\rlap{\lower3.7pt\hbox{$\mathchar\sim$}}
\raise1pt\hbox{$<$}\,}

\begin{document}

\title{Damping of supernova neutrino transitions\\
in stochastic shock-wave density profiles}
\author{Gianluigi Fogli, Eligio Lisi, Alessandro Mirizzi }
\address{ Dipartimento di Fisica
                and Sezione INFN di Bari\\
                Via Amendola 173, 70126 Bari, Italy\\}
 
\author{Daniele Montanino}
\address{Dipartimento di Fisica
                and Sezione INFN di Lecce\\       
                  Via Arnesano, 73100 Lecce, Italy\\}                
 
\date{3 March 2006}

\begin{abstract}
Supernova neutrino flavor transitions during the shock wave propagation
are known to encode relevant information not only   about the matter density profile
but also about unknown neutrino properties, such as the mass hierarchy 
(normal or inverted) and the mixing angle $\theta_{13}$. While previous studies
have focussed on ``deterministic'' density profiles, we investigate the effect
of possible stochastic matter density fluctuations in the wake of  
supernova shock waves.
In particular, we study the impact of small-scale fluctuations on the electron 
(anti)neutrino survival probability, and on the observable 
spectra of inverse-beta-decay events in future water-Cherenkov 
detectors. We find that such fluctuations, even with relatively small amplitudes,
can have significant damping effects on the flavor transition pattern,
and can partly erase the shock-wave imprint on the observable time spectra,
especially  for $\sin^2 \theta_{13} \gtrsim {\mathcal O}(10^{-3})$.
 \end{abstract}


\maketitle

\section{Introduction} 

Future observations of supernova (SN) neutrinos in underground detectors
represent a subject of intensive investigation in astroparticle physics.
In this context, matter effects associated to neutrino flavor transitions 
in the
SN envelope have been widely studied as a unique tool to probe, at the same
time, neutrino properties and supernova astrophysics 
(see, e.g.\ \cite{review,Raffrev,Digherev,Cavanna} for recent reviews).
In particular, the effects of supernova shock waves
on neutrino flavor transitions in matter
\cite{Schi} are gaining increasing attention in the recent 
literature~\cite{Taka,Luna,Foglish,Tomas,KaWa,Dasgupta}.
Indeed, for a few seconds after the SN core bounce, the
strong time dependence of the shock-wave profile can
leave peculiar signatures on the time structure of the neutrino 
events, which could be monitored in 
large, real-time detectors~\cite{FogliMega,Barger:2005it,Choubey:2006aq}.

So far, studies of supernova neutrino flavor transitions during the 
shock-wave propagation have been based on ``deterministic''
matter density profiles, assumed to be known (or at least knowable, 
in principle) 
both in time and in radial dependence. However, 
stochastic density fluctuations and inhomogeneities, of various 
magnitudes and correlation lengths, may reasonably arise in the 
wake of a shock front.  
Possible causes of these inhomogeneities include microscopic fluctuations
in the nascent neutron star~\cite{Keilconv} and large-scale fluctuations
between the proto-neutron star and the supernova envelope due to hydrodynamical
instabilities~\cite{Kifon,Bura05,Scheck:2006rw}. At early times 
($\lesssim 1$~s after bounce), 
post-shock convection overturns can also produce large density anisotropies.
A supernova neutrino ``beam'' traveling to the Earth might thus experience 
stochastic matter effects while traversing the stellar envelope.

Concerning neutrino oscillations, the phenomenology of possible stochastic
matter density fluctuations has been investigated in several contexts, 
with emphasis on general 
properties~\cite{Sawyer,Loreti:1994ry,Balan96,Benatti} and on the solution
to the solar neutrino 
problem~\cite{Nuno96,Burg96,Bykov,Burg02,Bal03,Guzzo}. Supernova neutrinos 
have specifically been considered in relatively few cases~\cite{LoreSN}, and 
only for a \emph{static\/} profile. In general, it is found 
that the typical effect of
random fluctuations on neutrino oscillations is to wash out the 
phase information (if any) and to damp the flavor transition pattern
in the energy domain. 
It seems thus worthwhile to revisit this topic
in the context of \emph{dynamic\/} SN density profiles, 
which offer a complementary handle in the time domain.%
\footnote{This possibility was mentioned in passing in the seminal Ref.~\cite{Schi}.}
Moreover, stochastic density fluctuations are expected to arise more easily
in the wake of a SN shock wave than in a relatively static stellar environment.

In this work we try to explore quantitatively the entangled effects
of shock waves and of possible stochastic fluctuations (behind the shock front) 
on supernova neutrino flavor 
transition probabilities and observables. Unfortunately, current
core-collapse SN simulations 
do not yet offer a clear input for such phenomena,
both because the details of the explosion mechanism are not well understood yet, and
because current computer resources do not allow to resolve density
variations at scales smaller than, say, $O(10)$ km. Therefore, 
some assumptions and simplifications are unavoidable. In particular,
we shall limit ourselves to small-scale and small-amplitude fluctuations,
which are definitely not excluded by current
simulations, and which allow a simple perturbative approach.

Our work is organized as follows.
In Section~2 we introduce the notation for the neutrino mass and mixing parameters 
and for the electron
(anti)neutrino survival
probability $P_{ee}$ characterizing supernova
neutrino transitions.  
In Section~3 we parametrize the ``fluctuating'' shock-wave profile and  discuss an
approximate analytic expression for the survival probability
$P_{ee}$,  which neatly includes the damping effect induced by 
density fluctuations. In particular, the analytical derivation of $P_{ee}$ from the
neutrino master equation
for the density matrix, together with a comparison with representative numerical solutions,
are given in the Appendix.
We find that,
for $\sin^2 \theta_{13} \gtrsim {\mathcal O}(10^{-3})$, 
small-scale stochastic fluctuations 
can suppress the imprint of the shock wave
on the flavor transition pattern in the time domain.
In Section~4 we discuss
a specific experimental application for the case of antineutrinos in 
inverted hierarchy, by calculating the effects of fluctuations on
positron event spectra
observable in water-Cherenkov detectors through inverse beta decay. 
Conclusions and prospects for further developments are given in Section~5.
The bottom line of our work is that some ``smearing'' of  
shock-wave signatures in neutrino observables is 
to be expected, if stochastic matter fluctuations are present in the
wake of the shock front. Improvements in supernova hydrodynamical simulations
will greatly help to clearify this issue.

\section{Three-neutrino mixing framework}
\label{nota}
In this section we set the notation for three-neutrino mixing
and for the survival probability of electron (anti)neutrinos $P_{ee}$. 
In the current ``standard'' $3\nu$ scenario, 
the dominant 
parameters relevant to flavor
transitions in supernovae are the largest 
squared mass difference $\Delta m^2$
and the mixing angles $\theta_{12}$ and $\theta_{13}$ (see, 
e.g., \cite{Dighe,LisiSN}). In numerical examples, we take as reference 
values (close to their best fits \cite{Fogl2005})
\begin{eqnarray}
\Delta m^2 &\simeq & 2.4\,
\times 10^{-3}\mathrm{\ eV}^2\ ,\\
\sin^2\theta_{12} &\simeq & 0.3 \,\ .
\end{eqnarray}
The sign of $\Delta m^2$ distinguishes the cases of  normal hierarchy  
 (NH: $+\Delta m^2$) and  inverted hierarchy (IH: $-\Delta m^2$). 
Concerning the mixing parameter $\sin^2 \theta_{13}$, we shall
use representative values below the current
upper limits ($\sin^2 \theta_{13} < \mathrm{few}\times 10^{-2}$~\cite{CHOOZ}).

The smallest squared mass difference 
($\delta m^2\simeq 8\times 10^{-5}$ eV$^2$~\cite{Fogl2005}) 
is such that $\delta m^2/\Delta m^2\ll 1$; together with the
smallness of $\sin^2\theta_{13}$, this fact guarantees, to a very good
approximation, the factorization of the $3\nu$ dynamics
into a ``low'' ($L$) and a ``high'' ($H$) $2\nu$  subsystem. In other words,
the electron (anti)neutrino survival probability $P_{ee}$, up to
small terms of $O(\sin^2\theta_{13},\delta m^2/\Delta m^2)$,
can be expressed as
\begin{equation}
P_{ee} \simeq  P_{ee}^{L} \cdot  P_{ee}^H \,\ ,
\end{equation}    
where $P_{ee}^L$ and $P_{ee}^H$ represent  effective 
electron (anti)neutrino survival probabilities in the $2\nu$ 
subsystems (see
\cite{KuPa,Dighe,LisiSN,KuoSN,MikSmir} and references therein). 

Neglecting Earth matter effects (not included in this work),
one has~\cite{LisiSN,Dighe,Foglish}
\begin{equation}
\label{casesL} P_{ee}^L \simeq  \left\{
\begin{array}{ll}
 \sin^2\theta_{12}  & (\textrm{for}\;\nu,\;\textrm{any}\;\ \textrm{hierarchy}), \\
 \cos^2\theta_{12}       & (\textrm{for}\;\overline\nu,\;\textrm{any}\;\ \textrm{hierarchy}), \\
\end{array}\right.
\end{equation}
and
\begin{equation}
\label{casesH} P_{ee}^H \simeq  \left\{
\begin{array}{ll}
 P_+^H  & (\textrm{for}\;\nu\;\textrm{in}\;\textrm{NH}\;\textrm{or}\;\overline{\nu}\;\textrm{in}\; \textrm{IH}), \\
 P_-^H  & (\textrm{for}\;\overline{\nu}\;\textrm{in}\;\textrm{NH}\;\textrm{or}\;\nu\;\textrm{in}\;\textrm{IH}), \\
\end{array}\right.
\end{equation}
where $P_{\pm}^H$ are defined below.

In general, $P_{ee}^{H}$ depends on $\theta_{13}$,
on the hierarchy, and
 on the
wavenumber $k_H$ in the $H$ subsystem,
\begin{equation}
\label{kH} \pm k_H = \pm \Delta m^2/2E\ \;\ (+\;\textrm{for\ NH} ,\, 
-\;\textrm{for\ IH}),
\end{equation}
and on the (anti)neutrino potential in matter \cite{Matt},
\begin{equation}
\label{V} \pm V(x)= \pm \sqrt{2}\, G_F\, N_e(x)\  \;\ (+\;\textrm{for}\;\nu ,
 -\;\textrm{for}\;\overline{\nu})
\end{equation}
where $N_e$ is the electron density at the supernova radius $x$.

Strong matter effects are generally expected when the ``level crossing
condition'' 
\begin{equation} 
\pm k_H \simeq \pm V(x_c)\ \;\ 
(\textrm{for}\;\nu\;\textrm{in}\;\textrm{NH}\;\textrm{or}\;\overline{\nu}\;\textrm{in}\;\textrm{IH}) ,
 \end{equation}
 is satisfied at (more than) one point $x_c$ (see \cite{LisiSN}). The crossing condition
requires equal signs for $V$ and $k_H$, so
it is not realized for $\overline\nu$
 in NH or $\nu$ in IH. 
 The physics of matter effects can be encoded in terms of the
 so-called crossing 
probability $P_c$~\cite{Dighe,Petcov}
\begin{equation}
\label{pcross} P_c =  \left\{
\begin{array}{ll}
 P_c(k_H,\sin^2 \theta_{13},V)  & (\textrm{for}\;\nu\;\textrm{in}\;\textrm{NH}\;\textrm{or}
\;\overline{\nu}\;\textrm{in}\;\textrm{IH}), \\
 \sim 0       & (\textrm{for}\;\overline{\nu}\;\textrm{in}\;\textrm{NH}\;\textrm{or}
\;\nu\;\textrm{in}\;\textrm{IH}), \\
\end{array}\right.
\end{equation}
where  $P_c \neq 0$ (=0) defines the case of nonadiabatic (adiabatic)
matter transitions.%
\footnote{Concerning the early literature on two-level crossings, it is amusing
to note that the classic Landau-Zener-St{\"u}ckelberg results 
for $P_c$~\cite{landau,zener,stuck} were
similarly obtained by Majorana~\cite{majo} in the context of spin-flip transitions in variable
magnetic fields, see the discussion in~\cite{digiac}. Curiously, this contribution by Majorana (well known in atomic
physics) is largely ignored in the neutrino literature.}

In the standard case (with no stochastic fluctuations)
the survival probability $P_{\pm}^H$ at the Earth (averaged over many
oscillation cycles) is related
to the  crossing probability $P_c$ through the
well-known Parke's formula~\cite{Parke},
\begin{equation}
P^H_{\pm} \simeq  \frac{1}{2} + \left(\frac{1}{2} -P_c \right) \cos 2 \theta_{13}
\cos 2 \tilde{\theta}_{13}^{\pm} (x_0) \ \ (\mathrm{no\ fluctuations})\ ,
\label{parke}
\end{equation}
where $x_0$ is the neutrino production point, and
\begin{equation}
\cos 2 \tilde{\theta}_{13}^{\pm} (x) = 
\frac{\cos 2 \theta_{13} \mp V(x)/k_H}
{\sqrt{(\cos 2 \theta_{13} \mp V(x)/k_H)^2 + (\sin 2 \theta_{13})^2}} \,\ ,
\label{eq:cosmatt}
\end{equation}
defines the mixing angle $\tilde{\theta}_{13}^{\pm}$ in  matter. 
In Eq.~(\ref{eq:cosmatt}), the upper and lower signs correspond to the upper
and lower signs of $P^H_\pm$ in Eq.~(\ref{casesH}). Due to the
high matter density at the origin ($|V(x_0)/k_H|\gg 1$), in Eq.~(\ref{parke}) one can
simply put
\begin{equation}
\label{cos1}\cos 2 \tilde{\theta}_{13}^{\pm}(x_0) \simeq \mp 1 \,\ .
\end{equation}

A final remark is in order.
In the presence of density fluctuations with small amplitude, one does
not expect that the crossing probability $P_c$
is significantly perturbed, being related to a ``local'' nonadiabatic effect. The
relation between $P^H_\pm$ and $P_c$ in Eq.~(\ref{parke})
is instead expected to change, being related
to the ``global''  propagation history within the supernova matter.
This intuitive picture, which enters in the analytical generalization
of Eq.~(\ref{parke}) discussed in the next section, is confirmed by numerical
calculations (see the Appendix). Similarly, one does not expect that
 small-amplitude fluctuations can ruin the effective $L\otimes H$ factorization,
which relies only on the smallness of $\theta_{13}$ and of $\delta m^2/\Delta m^2$.
This expectation is also supported by the smallness of 
fluctuation effects within the $L$ subsystem in our framework (see the last
paragraph in Sec.~3.3).  
A full numerical confirmation of the $L\otimes H$ factorization in the
presence of stochastic effects, however, would involve
rather heavy calculations of exact $3\nu$ solutions, which are beyond the scope
of
this work.

\section{Effects of stochastic matter fluctuations on SN neutrino transitions}
\label{randeq}

In this Section we parametrize the SN stochastic density fluctuations 
through some simplifying assumptions (Sec.~\ref{fluc}),
characterize their effects through
an analytical presciption for the calculation of 
the probability $P_{\pm}^H$ (Sec.~\ref{3nu}), and discuss the time dependence 
of $P_{\pm}^H$ in the presence of fluctuations 
with increasing amplitude (Sec.~\ref{analyP}).

\subsection{Shock-wave density profile and stochastic fluctuations}
\label{fluc}

We make the reasonable assumption that stochastic fluctuations arise 
only {\em after\/} the passage of the shock-wave.
Behind the shock front, fluctuations can be described as fractional (random)
variations $\xi(x)$ of an ``average'' neutrino potential $V_0$,
\begin{equation}
V(x) = V_0(x) + \delta V(x) = V_0(x) [1+ \xi (x) ] \,\ ,
\label{eq:fluct}
\end{equation}
with vanishing mean 
   $\langle \xi(x) \rangle =0$, and with variance
 $\xi^2=\langle \xi(x)^2 \rangle$. In the absence of
further information, we simply assume that $\xi$ is both constant and small
[$O(\mathrm{few} \%)$].%
\footnote{In a different context (Sun density profile), the combination
of all available solar and KamLAND data sets an upper 
limit $\xi<5\%$ at 70 $\%$ CL  for the solar density fluctuations, 
assuming a fluctuation length scale $L_0=10$~km~\cite{Bal03}.}

We also assume that the fluctuations arise at relatively small length scales
$L_0$, as compared with the neutrino oscillation length in matter $\lambda_m$.
In the region where matter effects are relevant ($V \simeq k_H$),
this condition reads:
\begin{eqnarray}
L_0 \ll \lambda_m & \simeq &
 \frac{2 \pi }{k_H \sin 2 \theta_{13}}  \nonumber \\
  &\simeq & 
60 \,\ \textrm{km} \left(\frac{E}{10 \,\ \textrm{MeV}} \right)
\left( \frac{2 \times 10^{-3} \,\ \textrm{eV}^2}{\Delta m^2} \right)
\left(\frac{0.2}{\sin 2 \theta_{13}} \right)  \,\ .
\label{smallam}
\end{eqnarray}
In all numerical examples we shall assume a representative value 
\begin{equation}
L_0 \simeq 10 \,\textrm{km} \,\ .
\label{length}
\end{equation}
 Notice that this scale
is of the same order as the radial hydrodynamical fluctuations 
described in~\cite{Keilconv},
as well as of the neutrinosphere radius~\cite{review}, which sets the transverse size of
the SN neutrino beam in the Earth direction.

The small-scale assumption in Eqs.~(\ref{smallam}) and~(\ref{length})
 implies that the autocorrelation function
of the random field $\xi(x)$ can be written as an effective delta-function 
\cite{Loreti:1994ry,Balan96,Benatti,Nuno96}
\begin{equation}
\label{delta}
\langle \xi(x_1) \xi(x_2) \rangle = 2 L_0 
 \xi^2  \delta(x_1-x_2) \,\ .
\end{equation}
In a nutshell, fluctuations between points much closer (much farther)
than one neutrino oscillation length in matter are assumed to be
fully correlated (totally
uncorrelated). Given Eq.~(\ref{delta}), the master equation for the neutrino
flavor evolution is determined (see the Appendix).

\begin{figure}[t]
\vspace*{-0.0cm}\hspace*{-0.2cm}
\begin{center}
\includegraphics[scale=0.55]{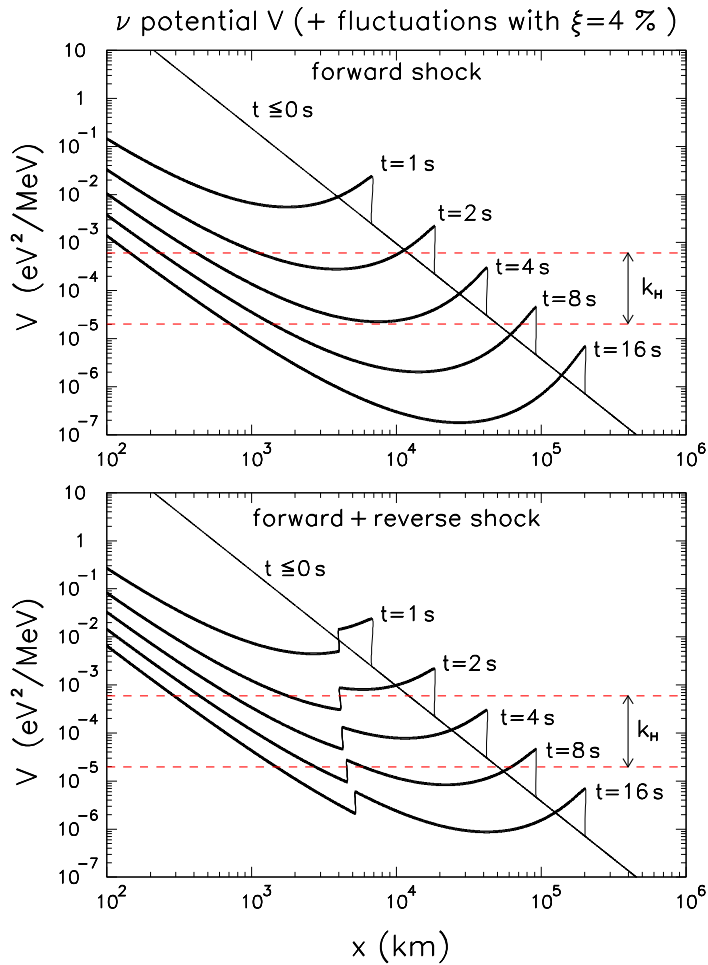}
\vspace*{+0.2cm} \caption{\label{pot}
\footnotesize\baselineskip=4mm Simplified
radial profiles of the neutrino potential $V(x)$ at different 
post-bounce times $t$ (1, 2, 4, 8, and 16 s). The thicker curves behind the shock front represent fluctuations
of amplitude $\xi \leq   4 \%$ of the local matter potential $V$. 
Upper panel: Forward shock
only. Lower panel: Forward plus 
reverse shock. In both cases, the static (pre-bounce) profile ($t\leq 0$~s) is
also shown. The band within dashed lines marks the region where  
SN matter effects are potentially 
important ($V\simeq k_H$, with $E=2$--60~MeV).}
\end{center}
\end{figure}

Figure 1 shows the 
(fluctuating) shock-wave  profiles used in this work.
We take the unperturbated shock-wave  potentials  $V_0(x)$
from our previous semplified parametrizations in~\cite{Foglish,FogliMega}.
In the region behind the shock front, the 
thicker curves represent a band of fluctuations with fractional 
deviation $\xi \leq  4 \%$. 
The upper panel refers to the neutrino potential $V$ in the presence of the forward
shock only. 
 In the lower panel we have
taken into  account also a possible reverse shock,
characterized by a smaller density jump at the front~\cite{Tomas}.
In both panels, 
we also show the band spanned by the neutrino wavenumber $k_H=\Delta m^2/2E$
for $E\in[2,60]$ MeV. Notice that the (crossing) condition for large matter effects 
($V\simeq k_H$) is basically unperturbed by fluctuations as small as those
in Fig.~1.

\subsection{Analytical results}
\label{3nu}

As anticipated in Sec.~\ref{nota}, the survival probability $P^H_{\pm}$ 
in the $H$ subsytem is modified by matter density fluctuations. 
In general, the calculation of $P^H_{\pm}$ requires that 
the stochastic master equation (see Appendix) is solved numerically. Nevertheless,
as shown in \cite{Burg96} and described in detail in the Appendix, 
small-scale and small-amplitude fluctuations allow to use
perturbative techniques which, already at first order, 
provide rather accurate analytical
approximations. In the context of SN neutrinos, 
the analytical prescription reads
\begin{equation}
P^H_{\pm} \simeq  \frac{1}{2} \mp \left(\frac{1}{2} -P_c \right) e^{-\Gamma_{\pm}} 
\cos 2 \theta_{13}  \,\ , 
\label{eq:prodamp}
\end{equation}
which generalizes \cite{Burg96} 
Parke's formula (\ref{parke}) [in the conditions of Eq.~(\ref{cos1})] through a simple 
exponential damping factor $e^{-\Gamma_{\pm}}$, whose exponent
is given by
\begin{equation}
\Gamma_{\pm} =  \int_ {x_0}^{x_s}  {\mathcal D}(x) \,\  \sin^2 2 \tilde{\theta}_{13}^{\pm} (x)  dx \,\ ,
\label{gamma}
\end{equation}
where
\begin{equation}
\label{dampingpar} {\mathcal D}(x) = \xi^2  V_0^2(x) L_0 \,\ ,
 \end{equation}
and the mixing angle in matter $(\tilde\theta_{13}^\pm)$ is defined as in Eq.~(\ref{eq:cosmatt}).
The domain of the damping integral $\Gamma_{\pm}$ is  
 the region behind the forward shock front position $x_s$,
where stochastic fluctuations are assumed to arise.  
For strong damping ($\Gamma_{\pm} \gg 1$) one gets
the limit $P^H_{\pm} \to 1/2$, corresponding to a sort of complete 
``flavor depolarization,'' where the two effective $\nu$ states in the $H$ subsystem
are democratically mixed.

All stochastic matter effects are embedded in the damping factor 
$e^{-\Gamma_{\pm}}$, while the crossing probability $P_c=P_c(k_H,\sin^2 \theta_{13},V(x))$ in Eq.~(\ref{eq:prodamp}) is basically unperturbed.%
\footnote{This fact is also verified a posteriori through the 
very good agreement of the analytical recipe in Eq.~(\ref{eq:prodamp}) 
with the numerical solution, as shown in the Appendix. The check is notrivial,
since the accuracy of the analytical 
approximation in Eq.~(\ref{eq:prodamp}), discussed in~\cite{Burg96} 
in the context of the (static and monotonic) solar matter profile, cannot be taken for granted
in the (dynamic and nonmonotic) supernova profile.} 
In particular, 
$P_c$ can be evaluated analytically by considering the ordered 
product of crossing matrices along the average
(nonmonotonic) supernova shock-wave  density 
profile $V_0(x)$, as previously discussed
in~\cite{Foglish,FogliMega}. Therefore, apart from the numerical evaluation
of the integral in Eq.~(\ref{gamma}), $P^H_\pm$ can be computed
through convenient analytical approximations.

\subsection{Analysis of the survival probability $P^H_{\pm}$ in the
$H$ subsystem, and  comments on $P^L_{ee}$ in the $L$ subsystem}
\label{analyP}

In this Section we describe the behaviour of  $P^H_{\pm}$ as a function of time,
in the presence of
stochastic fluctuations with increasing amplitude $\xi$. The case of
no fluctuations is recovered for $\xi=0$.

\begin{figure}[t]
\vspace*{-0.0cm}\hspace*{-0.2cm}
\begin{center}
\includegraphics[scale=0.7]{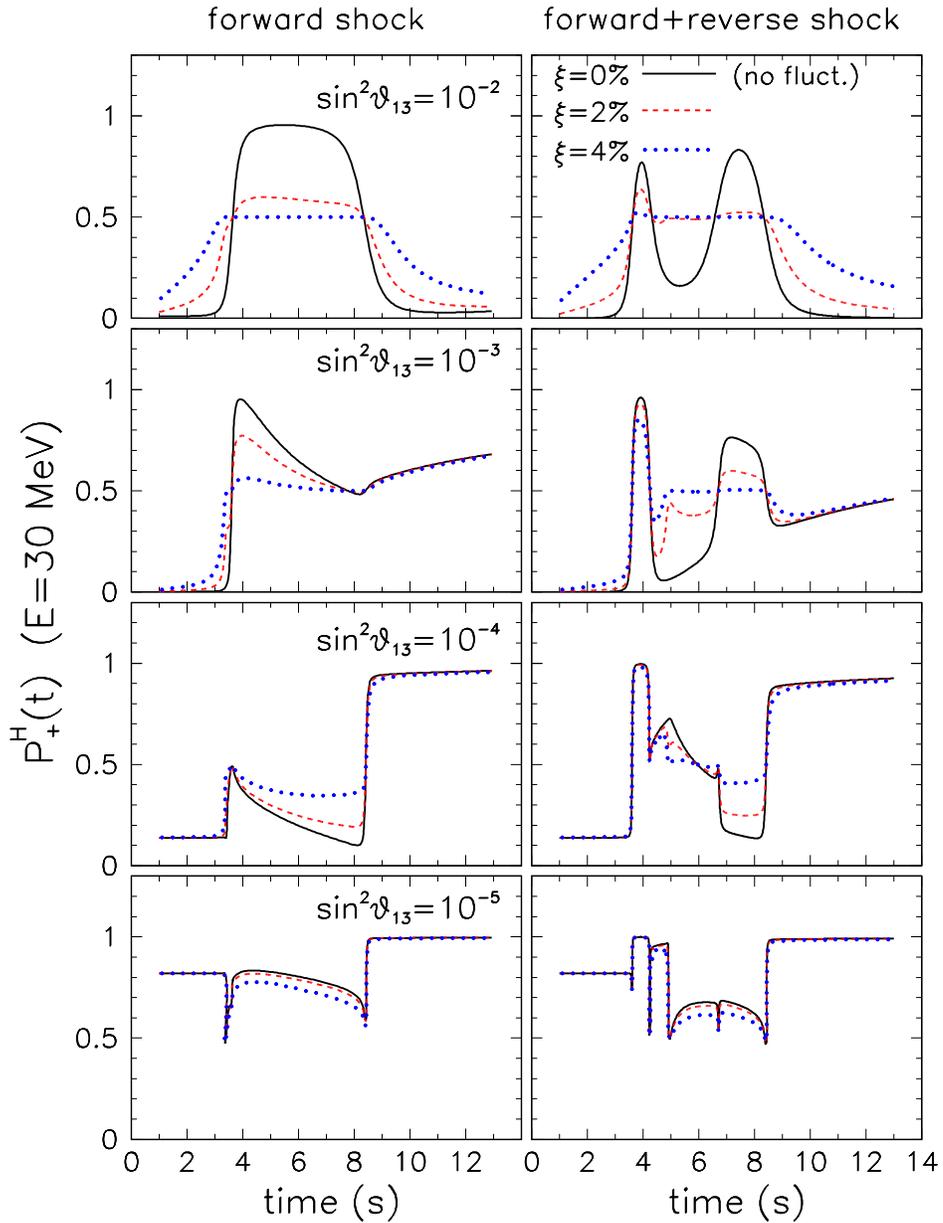}
\vspace*{+0.2cm} \caption{\label{lev} Probability function $P^H_{+}(t)$ (relevant
for $\nu$ in NH or $\overline\nu$ in IH)
 at $E=30$~MeV,  for four
representative values of $\sin^2{\theta_{13}}$, in the presence of density fluctuations
with fractional amplitude $\xi$  equal to $0 \%$ (solid curves), $2 \%$ (dashed curves) and
$4 \%$ (dotted curves). Left panels: forward shock
only. Right panels: forward plus reverse shock.
\footnotesize\baselineskip=4mm }
\end{center}
\end{figure}

Figure~\ref{lev} shows the variations of the survival probability $P_+^H$ 
(relevant for $\nu$ 
in NH or $\overline\nu$ in IH),
\begin{equation}
P_{+}^H \simeq \frac{1}{2} - \left(\frac{1}{2} -P_c \right) e^{-\Gamma_{+}} 
\cos 2 \theta_{13}  \,\ , 
\end{equation}
in the time interval 
$t \in [1,13]$~s,  for $\xi=0$
 (solid curves), $\xi=2\%$ (dashed curves) and  $\xi=4\%$ (dotted curves). The neutrino 
energy is fixed at $E=30$~MeV,  while $\sin^2 \theta_{13}$ ranges from 
$10^{-2}$ (upper panels) to $10^{-5}$ (lower panels). We include both the
case of forward shock only (left panels) and of forward
plus reverse shock (right panels).

In the absence of fluctuations, the function $P^H_{+}(t)$ in Fig.~2 shows strong, 
nonmonotonic variations: A clear signature 
of the strongly nonadiabatic flavor transitions
along the shock front, operative in the time window $t\in \sim[3,9]$~s (as discussed 
at length in~\cite{Foglish,FogliMega}). In the presence  
of stochastic fluctuations behind the shock front, however, the variations
are partly suppressed, as a result of the  ``flavor depolarization'' effect. 
As the amplitude $\xi$ increases,  
the survival probability $P^H_{+}$ gets closer to the saturation value
$1/2$. The effect of fluctuations is relevant in the same time window where
shock effects are operative, since the occurrence of the 
condition $V \simeq k_H$ (in multiple points $x_c$) leads to 
$\sin^2 2 \tilde{\theta}_{13}^+ \simeq 1$ and thus to  
a ``large'' damping factor [see Eq.~(\ref{gamma})] around $x_c$.
At earlier or later times, the enhancement of $\theta_{13}$ in matter is
more modest and so are fluctuation effects. 

In Fig.~2, the smearing effect of fluctuations appears to be 
particularly dangerous for the identification of shock-wave signatures: For instance,
for $\xi=4\%$, the double-peak structure associated to the forward+reverse 
shock (right panels) may be washed out, becoming more similar to the case
of forward shock only (left panels).  
A ``confusion scenario'' might arise,  limiting
 the potential to monitor
the shock-wave propagation through  neutrino flavor transitions.

Figure~2 also shows that, in general, damping effects are suppressed at
small values of $\theta_{13}$ (say, $\sin^2\theta_{13}\lesssim 10^{-3}$).
This fact can be understood by expanding the damping integral [Eq.~(\ref{gamma})] around
the crossing points $x_c$, where the integrand is locally maximal. At first order in
$x-x_c$, and using the fact that $V\simeq k_H$ around $x_c$, one gets then
\begin{equation}
 \Gamma_+ \simeq  \pi  \sin 2\theta_{13}  \sum_{x_c}  {\mathcal D}(x_c)
 \left|\frac{d \ln V}{dx}\right|_{x_c}^{-1} \,\ ,
 \label{gammappr}
\end{equation} 
which shows that the damping factor is basically proportional to 
$\theta_{13}$. 
In conclusion, for values of $\sin^2 \theta_{13} \gtrsim {\mathcal O}(10^{-3})$, 
small-scale stochastic fluctuations with a fractional amplitude of a few percent
might significantly suppress shock-wave effects on the  electron
neutrino survival probability  $P^H_{+}$. For smaller values of 
$\sin^2\theta_{13}$, fluctuation effects appear to be
less dangerous, but unfortunately the overall effect of shock waves on
$P_{+}^H$ is also small. 

\begin{figure}[t]
\vspace*{-0.0cm}\hspace*{-0.2cm}
\begin{center}
\includegraphics[scale=0.7]{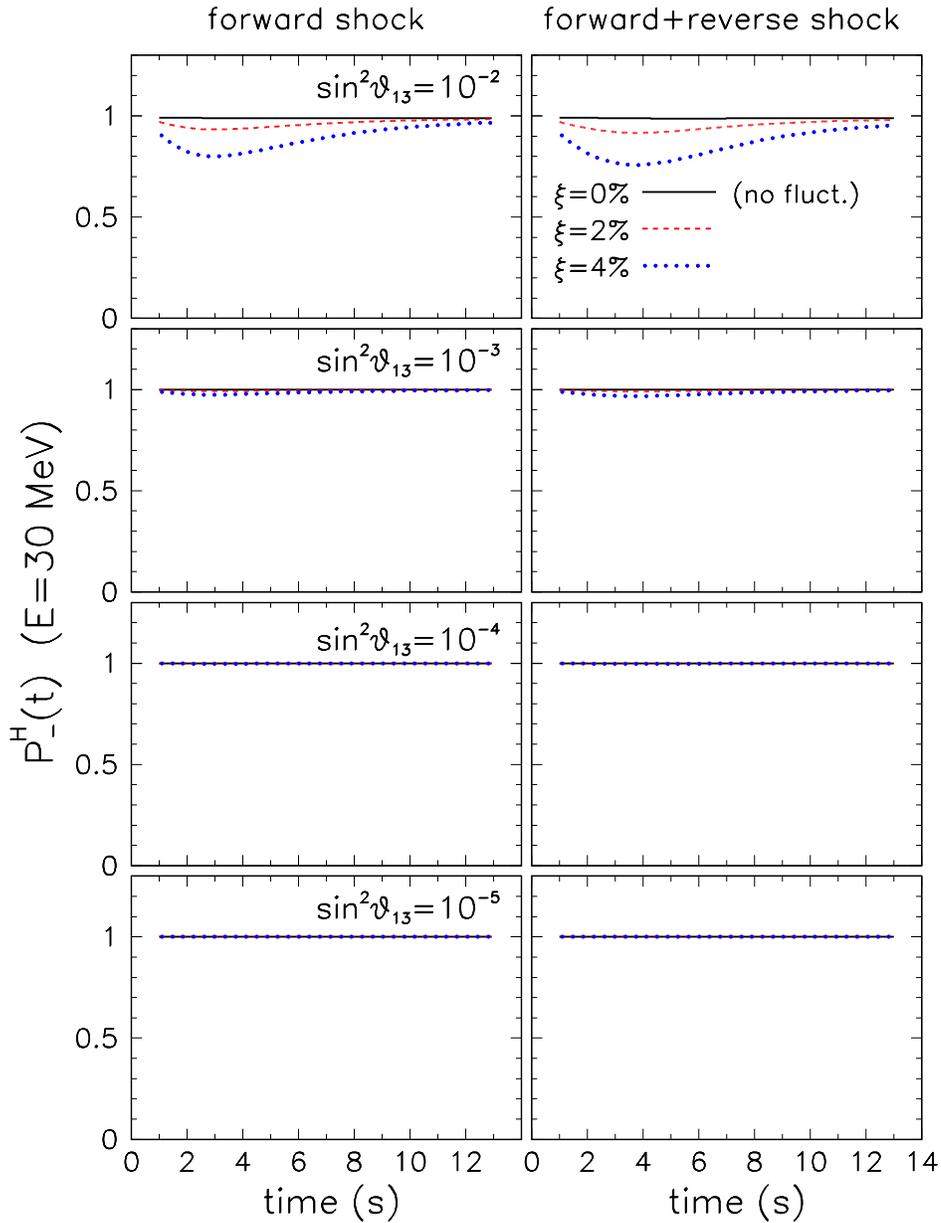}
\vspace*{+0.2cm} \caption{\label{levb} As in Fig.~2, but for the probability
function $P^H_{-}(t)$ (relevant
for $\nu$ in IH or $\overline\nu$ in NH).
\footnotesize\baselineskip=4mm }
\end{center}
\end{figure}

Figure~\ref{levb} shows the time evolution of the survival probability
$P^H_-$
(relevant for $\nu$ 
in IH or $\overline\nu$ in NH, where $P_c=0$),
\begin{equation}
P^H_- \simeq \frac{1}{2} \left( 1+ e^{-\Gamma_{-}} 
\cos 2 \theta_{13} \right) \,\ ,
\label{eq:phm} 
\end{equation} 
in the same format of Fig.~\ref{lev}.
 In the absence of matter fluctuations ($\xi=0$), it is 
 $\Gamma_-=0$ and $P^H_- \simeq 1$ trivially 
 (being $\cos2\theta_{13}\simeq 1$ in all cases shown in Fig.~3).
 In the presence of  fluctuations, the damping
 term $e^{-\Gamma_-}$ tends to lower $P^H_-$ (down to the plateau value
$1/2$  for large $\Gamma_-$).
However, the relative variations of the function $P_-^H$ 
in Fig.~\ref{levb} are not as strong as for $P_+^H$  in Fig.~2, since 
 the  integrand in $\Gamma_-$ is never ``resonant'', as it happens
instead for $\Gamma_+$.  For small $\theta_{13}$, it 
is simply 
  \begin{equation}
  \Gamma_- \simeq \sin^2 2 \theta_{13} \int_ {x_0}^{x_s} \frac{{\mathcal D}(x)}
  {(1 +V(x)/k_H)^2} \;\ ,
  \end{equation}
namely, the damping factor $\Gamma_-$  decreases with 
  $\sim \theta_{13}^2$, and its effect vanishes more rapidly than for $\Gamma_+$.
This fact explains why, in Fig.~3, fluctuation effects are visible
only at  larger values of $\theta_{13}$  (upper panels, $\sin^2\theta_{13}= 10^{-2})$
as compared with Fig.~2.
The absence of level crossing effects in $P_-^H$ strongly reduces the sensitivity
to the details of the
shock-wave profile, so that for $\xi>0$ there is hardly any difference
between the left and right upper panels in Fig.~3.
    
From the previous discussion, stochastic density fluctuations appear to be
most relevant  in the cases where $P_+^H$ (rather than $P_-^H$) is involved, i.e.,
  neutrinos
 in normal hierarchy or  antineutrinos in inverted hierarchy. 
 For this reason, in the next section we shall examine the impact of these results
in a phenomenologically relevant case ($\overline \nu$ in IH), by calculating observable spectra
of positrons induced by $\overline\nu_e$ through inverse beta decay. 

Finally, we comment about the effects of fluctuations on the survival probability
$P_{ee}^L$ in the ``low'' ($L$) subsystem. 
In the presence of density fluctuations, 
this probability can be calculated as
\begin{equation} 
P_{ee}^L \simeq \frac{1}{2} \mp \frac{1}{2} e^{- \Gamma_{\pm}(L)} \cos 2 \theta_{12} \,\ ,
\label{peelfl}
 \end{equation}
 which is analogous to Eq.~(\ref{eq:prodamp}) for the adiabatic case $P_c=0$
 (the adiabaticity of the $L$ subsystem is discussed
 in~\cite{Foglish}). The damping integral in the above equations reads
 \begin{equation}
\Gamma_{\pm}(L)= 
\int_ {x_0}^{x_s}  {\mathcal D}(x) \,\  \sin^2 2 \tilde{\theta}_{12}^{\pm} (x)  dx \,\ ,
\label{gammaL}
\end{equation}
where the definition of $\tilde{\theta}_{12}^{\pm}$ is analogous to
that in Eq.~(\ref{eq:cosmatt}),
but in terms of $\theta_{12}$ and of $k_L= \delta m^2 /2E$.
Since $k_L$ is a factor $\sim 30$ smaller than $k_H$, one has tipically $V/k_L \gg 1$
for $x < x_s$, and thus $\sin^2 2 \tilde{\theta}_{12}(x) \ll 1$ in Eq.~(\ref{gammaL}),
except possibly at late times ($t \gtrsim 10$~s, see Fig.~1), where the crossing
condition $V\simeq k_L$ can be realized, leading to $\sin^2 \tilde{\theta}_{12}\simeq 1$
for \emph{neutrinos}. 
However, at late times the neutrino luminosity is also small. Therefore, the damping
effects of $\Gamma_{\pm}(L)$ are either small or suppressed by a low luminosity, 
and can be safely neglected for the purposes of our work. One can simply take
$e^{-\Gamma_{\pm}(L)}\simeq 1$ in Eq.~(\ref{peelfl}), thus recovering
 Eq.~(\ref{casesL}).

\section{Observable positron spectra from SN $\overline{\nu}$
in inverted hierarchy}
\label{sec:posit}
 
 As an application of the previous results, we study the effects 
 of stochastic matter fluctuations on the observable  
 positron time spectra detectable through the inverse beta-decay reaction
 \begin{equation}
 {\overline\nu}_e + p \to n + e^+ \,\ ,
 \end{equation}
which represents the main SN neutrino detection channel in
 present~\cite{SKde} and planned
 Cherenkov detectors~\cite{UNNO,Jung,HK03,Vill}, as well as in 
scintillation detectors~\cite{Cadonati,Aglietta,ober}. 
Therefore, in this Section we focus on electron antineutrinos,
in the relevant case of inverted hierarchy,%
\footnote{An analogous study (not presented here) could
 be performed for the neutrino channel in normal hierarchy, in
 the context of $\nu_e$-sensitive detectors,
 such as  liquid argon time proportional chambers~\cite{Gil-Botella:2004bv}.}
where shock-wave 
(plus fluctuation) effects
become potentially observable through the $H$ subsystem dynamics,
\begin{equation}
P_{ee} \simeq  \cos^2\theta_{12} P_{+}^H \,\ .
\end{equation}

Positron 
event rates are calculated for a reference 0.4 Mton water-Cherenkov detector, 
with neutrino spectra at the supernova source, interaction cross sections,
and detection parameters fixed as in our previous work
\cite{FogliMega}, to which the reader is referred for further details.
Here we just mention that in \cite{FogliMega} two representative
 ``low'' and  ``high''
positron energy bins (at $E_{\rm pos}= 20 \pm 5$ and $45\pm 5$~MeV, respectively)
were identified as good relative tracers of shock-wave effects. Indeed, as discussed 
in~\cite{FogliMega} (for no fluctuations),
 the time spectra at relatively high energy ($E_{\rm pos}= 45 \pm 5$~MeV)
  show strong signatures of the shock-wave propagation, especially for increasing $\sin^2\theta_{13}$,
through nonmonotonic time variations of the event rates. 
Conversely, the shock-induced time variations in the bin
$E_{\rm pos}= 20 \pm 5$~MeV are significantly smaller; at such energy,
in fact, the initial electron and non-electron antineutrino fluxes 
(as used in our calculation input)
happen to be approximatively equal, and the effects
of flavor transitions largely cancel.

\begin{figure}[t]
\vspace*{-0.0cm}\hspace*{-0.2cm}
\begin{center}
\includegraphics[scale=0.7]{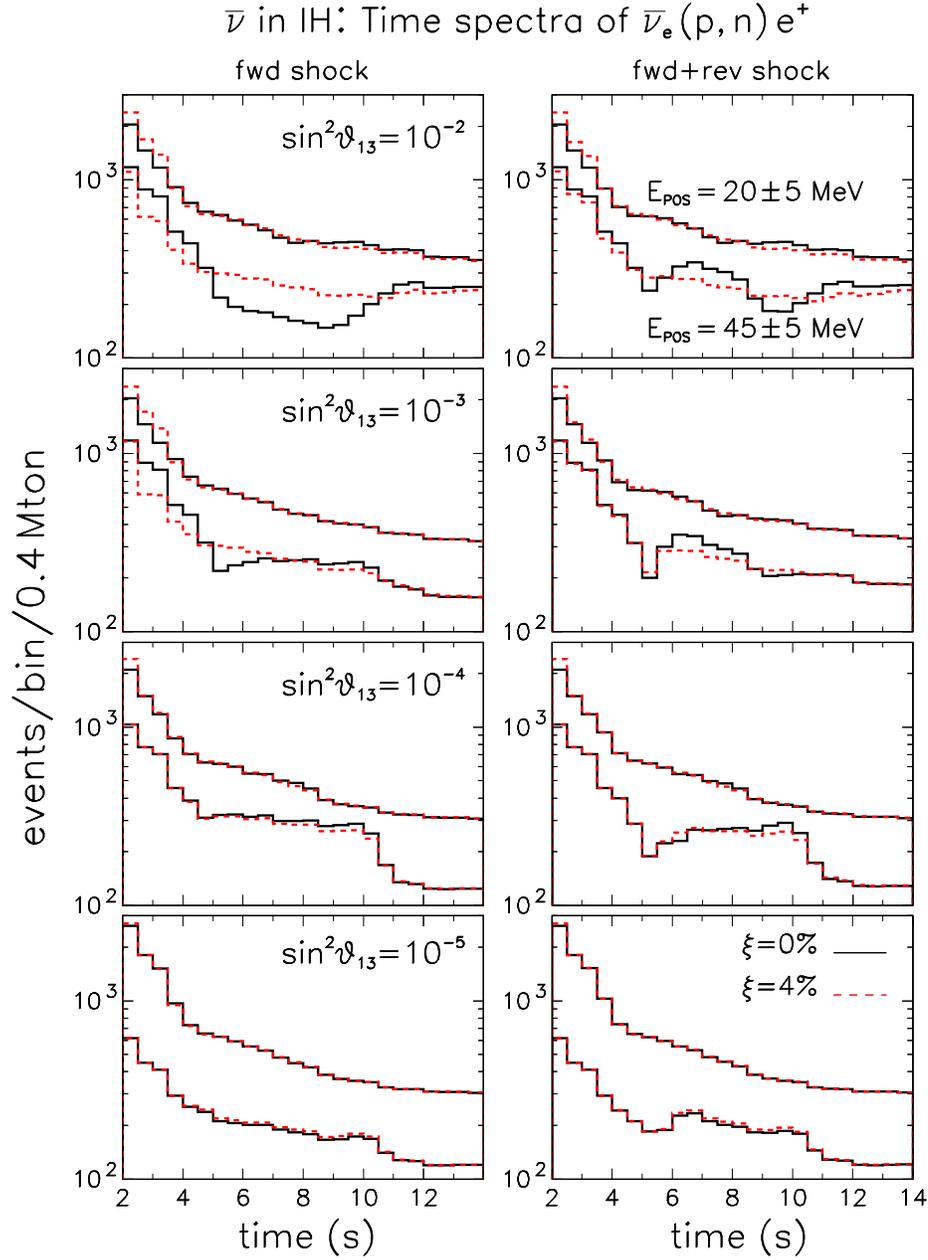}
\vspace*{+0.2cm} \caption{\label{rate} Absolute time spectra of positron events
induced by $\overline \nu_e$ in IH in a 0.4 Mton water-Cherenkov detector,  
in the presence of forward shock (left panels) and
forward plus reverse shock (right panels).  Four representative values of
$\sin^2 \theta_{13}$ are considered. The solid histograms refer to the case of
 no fluctuation ($\xi=0$), while the dashed ones refer to the case of fluctuations
with $\xi=4\%$. In each panel, the upper (lower) couple of histograms
 refers to the positron energy bin $E_{\rm pos}=
20 \pm 5$~MeV ($E_{\rm pos}=
45 \pm 5$~MeV).
\footnotesize\baselineskip=4mm }
\end{center}
\end{figure}

Figure~\ref{rate} shows absolute time spectra of events within
the two  reference positron energy bins
($E_{\rm pos}= 20 \pm 5$~MeV and $E_{\rm pos}= 45 \pm 5$~MeV) for 
four representative   values of $\sin^2 \theta_{13}$,
and for both $\xi=0$ (no fluctuations, solid histograms)
and $\xi=4\%$ (dashed histograms).
The left and right panels refer to the case of forward shock only
and of forward plus reverse shock, respectively.   
In the presence of stochastic density fluctuations ($\xi=4\%$ in
Fig.~\ref{rate}), the spectra for the low-energy positron bin are
basically unaffected, since all flavor transition effects (modified or
not by fluctuations) are small. The time spectra for the high-energy
positron bin ($E_{\rm pos}= 45 \pm 5$~MeV) are instead significantly smoothed
out by fluctuation effects for  
$\sin^2 \theta_{13} \gtrsim {\mathcal O}(10^{-3})$, as 
expected from the discussion of Fig.~2. Therefore, even at small amplitude
$(\xi=4\%)$, fluctuations can suppress the shock imprint on the time spectra, 
and make them qualitatively similar to those in normal 
hierarchy (where they are expected a priori to be smooth and monotonic).

\begin{figure}[t]
\vspace*{-0.0cm}\hspace*{-0.2cm}
\begin{center}
\includegraphics[scale=0.7]{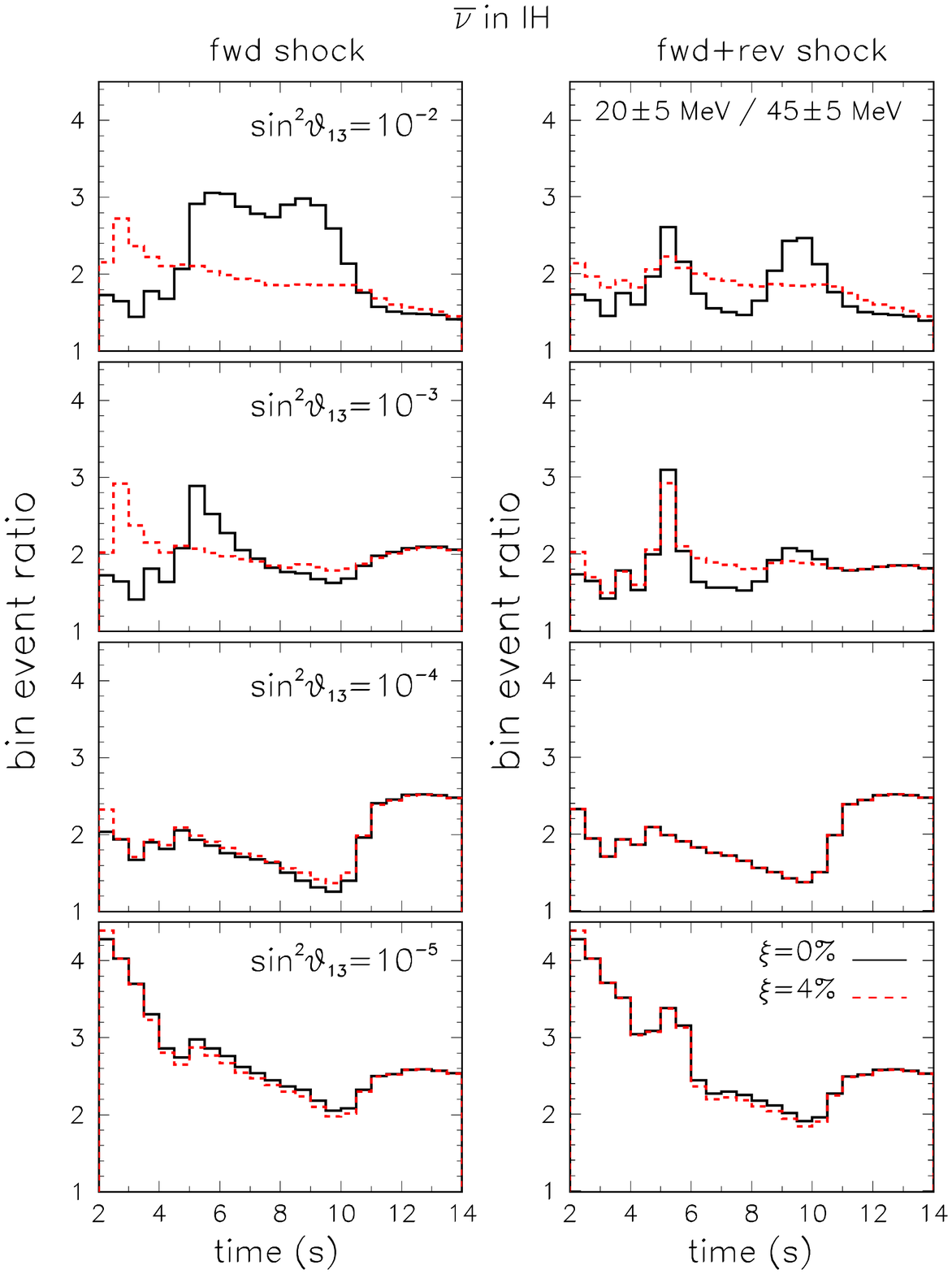}
\vspace*{+0.2cm} \caption{\label{ratio} 
Time dependence of the ratio of events between the
energy bins $E_{\rm pos}= 20 \pm 5$~MeV and 
$E_{\rm pos}= 45 \pm 5$~MeV  in the case of foward shock
only (left panels) and forward plus reverse shock (right panels).
The ratio refers to $\overline{\nu}_e$ in inverted hierarchy for
 no fluctuation ($\xi=0$, solid curves) and a fluctuation amplitude
$\xi=4\%$ of the local matter potential (dashed curves). 
\footnotesize\baselineskip=4mm }
\end{center}
\end{figure}

As suggested in \cite{FogliMega} (for the case of no fluctuations), the 
relative flavor transition effects at low and high energy are best
displayed by showing the ratio of the spectra at
$20\pm 5$ MeV and $45\pm 5$ MeV, where the first spectrum acts as a
``normalization'' factor.
Figure~\ref{ratio} shows such ratio  (in inverted hierarchy)
 for the same   values of $\xi$ and $\sin^2 \theta_{13}$
 of Fig.~\ref{rate}.    
 In the absence
of stochastic fluctuations ($\xi=0$), we recover the fact
 \cite{FogliMega} that the spectral ratio  can track the time variations 
 of the electron antineutrino survival  probability $P^H_{+}$, as evident
from a visual comparison of the solid
curves ($\xi=0$) in Figs.~2 and 5.%
\footnote{The analysis  in~\cite{FogliMega} actually refers to the
crossing probability $P_c$. However, in the absence of fluctuations,   
the smallness of $\theta_{13}$ implies that $P^H_{+}\approx P_c$,
see Eq.~(\ref{parke}).}
Unfortunately, stochastic density fluctuations (dashed curves with
$\xi=4~\%$ in Fig.~5) destroy this nice correspondence for  
 $\sin^2 \theta_{13} \gtrsim {\mathcal O}(10^{-3})$. The spectral ratio
becomes smoother, and the ``peaks and valleys'' induced by the
shock wave on $P_{+}^H$ disappear. 
The damping effect implies
a net loss of information about the shock wave in the observable neutrino spectra.

\section{Summary and conclusions}
\label{conclu}
  
Stochastic  density fluctuations in supernova matter may produce observable
effects on the supernova neutrino signal. In this context, previous works have 
 been focussed on static and monotonic profiles. In this paper, we have investigated the case of  
time-dependent and non-monotonic profiles, embedding forward (plus reverse) shock
propagation, as suggested by recent SN numerical simulations. 
In the hypothesis of 
small-scale ($L_0 \sim {\mathcal O} (10$~km)) and 
small-amplitude ($\xi \lesssim$ few$\%$) fluctuations,
we have discussed an analytical recipe to evaluate the SN electron (anti)neutrino survival probability
$P_{ee}$, which accounts for both  standard  matter transitions and 
additional ``flavor-depolarization'' effects induced by the fluctuations, 
and which accurately reproduces  the numerical 
results (see the Appendix).

We find that,  stochastic
fluctuations---possibly arising after the shock front passage---may significantly 
suppress the imprint of SN shock waves on $P_{ee}$ in the time domain,
the more the larger $\sin^2\theta_{13}$. For decreasing
$\sin^2 \theta_{13}$, fluctuation effects decrease, but shock-wave signatures
 also become less pronounced. 
An application  to  observable time spectra of positron events in large water-Cherenkov detectors
shows that, for the phenomenologically relevant case of inverted hierarchy,
the time spectra can easily loose any imprint of the shock wave
for $\sin^2 \theta_{13} \gtrsim {\mathcal O}(10^{-3})$. Therefore, in the
presence of stochastic fluctuations, it might be difficult to 
 ``monitor'' the shock-wave in real time in a future galactic
 SN explosion, or to find unmistakable signatures of inverted mass hierarchy effects.

In this context, future work might include a ``fluctuation analysis''  of the 
SN~1987A neutrino events~\cite{Mirizzi:2005tg} and of the expected 
supernova  relic neutrino spectrum~\cite{Malek:2002ns,Foglirelic,Yuksel:2005ae}.
For instance, some studies of the SN~1987A $\overline{\nu}_e$ in the case
of inverted hierarchy~\cite{Minakata:2000rx}  found a tension between data
and theory in the case of adiabatic
transitions (i.e., $\sin^2 \theta_{13} \gtrsim 10^{-4}$) and inverted hierarchy,
using a static SN profile.  The effect of both nonadiabatic crossings and of
damping behind the shock front might put these results in a different
perspective.
 
In any case, further studies will greatly benefit from improvements of numerical
SN simulations,
so as to follow the shock wave evolution with much higher space-time 
resolution than currently
possible. A better  theoretical understanding of the possibility and property of stochastic
density fluctuations in the SN envelope might also help to remove the simplifying assumptions
adopted in this work.

\section*{Acknowledgments}
 
This work is supported in part by 
the Italian ``Istituto Nazionale di Fisica Nucleare'' (INFN) and by the ``Ministero dell'Istruzione,  
Universit\`a e Ricerca'' (MIUR) through the ``Astroparticle Physics'' research project.

\appendix

\section{Stochastic neutrino master equation: analytical and numerical solutions}
 
For the sake of completeness,
in this Appendix we explicitly derive Eq.~(\ref{eq:prodamp}) 
(based on previous literature on the subject) and compare it
with our numerical results for representative SN shock-wave cases. 
To avoid an excessive ``$\pm$'' notation, we consider here 
only the case of neutrinos in normal 
hierarchy. The cases of antineutrinos and of inverted hierarchy are
 obtained by the replacements $V\to-V$ and $k_H\to-k_H$, respectively.

In  the presence of stochastic matter fluctuations, one must
consider the neutrino evolution within a statistical ensemble 
of density profiles. Averaging over the ensemble generally produces
a loss of the coherence in the system. In this context,
the appropriate formalism involves the neutrino density matrix $\rho$.

For a $2 \nu$ mixing problem (as in the $H$ subsystem), the density matrix in the
flavor basis $(\nu_e,\nu_a)^T$, where $\nu_a$ is any linear combination
of $\nu_\mu$ and $\nu_\tau$,
can be expressed
in terms of a ``polarization vector'' ${\bf P}$~\cite{MS86,Leo,Kimb,Duan}:
 \begin{equation}
 \rho = \frac{1}{2}(1 + {\bf P}\cdot \mathbf{\sigma}) \,\ ,
 \label{eq:dens}
 \end{equation}
where 
\begin{equation}
{\bf P} = 
\left(
\begin{array}{c}
2 \textrm{Re}(\nu_e^\ast \nu_a) \\
2 \textrm{Im}(\nu_e^\ast \nu_a) \\
2 |\nu_e|^2 -1
\end{array}
\right)
\label{eq:polar} \,\ ,
\end{equation}
and  $\sigma$ is the vector of Pauli matrices.
The length $|{\bf P}|$ of the polarization vector ($|{\bf P}|^2= \textrm{Tr} \rho^2$)
measures the degree of coherence of the system: $|{\bf P}| 
=1$ corresponds to a pure state, 
$0<|{\bf P}|<1$ to a partially mixed state, and $|{\bf P}|=0$ to a completely mixed 
state.

With or without fluctuations, the Hamiltonian for  neutrinos propagating in 
matter in the ``high'' subsystem is the usual one~\cite{Matt}:
\begin{equation}
H \equiv  \frac{1}{2} {\bf h} \cdot {\bf \sigma}= 
\frac{1}{2}\left[ k_H \sin 2 \theta_{13} \right] \sigma_1 + 
\frac{1}{2}
\left[ V(x) - k_H \cos 2 \theta_{13} 
  \right] \sigma_3  \,\ ,
\label{eq:Hamil}
\end{equation}
with
\begin{equation}
{\bf h} = (k_H \sin 2 \theta_{13} ,\,\  0,\,\  V(x) - k_H \cos 2 \theta_{13} ) \,\ .
 \label{eq:magn}
\end{equation}
In the presence of stochastic matter density fluctuations, 
the potential has been expressed as $V(x)=V_0(x)[1+ \xi(x)]$, with
$\delta$-correlated Gaussian fluctuations (see Sec.~3.1). In this case, 
the Liouville evolution of the fluctuations-averaged density matrix reads 
\cite{Loreti:1994ry,Balan96,Benatti}%
\footnote{For neutrinos propagating in a stochastic matter, the density
matrix [Eq.~(\ref{eq:dens})] and the polarization vector [Eq.~(\ref{eq:polar})]
must be understood as averaged over the ensamble of density profiles.}
\begin{equation}
 \frac{d}{d x}{\rho} = -i[H_0,\rho] - \frac{1}{4}{\mathcal D} [\sigma_3,[\sigma_3,\rho]] \,\ ,
 \label{eq:decoh}
 \end{equation}
where $H_0$ is the unperturbated Hamiltonian (for $\xi=0$) and ${\mathcal D}$ is the
same damping parameter as in Eq.~(\ref{dampingpar}).  

Equation~(\ref{eq:decoh}) has the well-known Lindblad form for dissipative
systems~\cite{Lindblad:1975ef}, where the first term in the right-hand side reproduces 
the standard Liouville equation for the $\nu$ density matrix, while the
second term  violates the conservation of $\textrm{Tr}\rho^2$
and allows transitions from pure to mixed states. 
 In terms of the polarization vector, 
 Eq.~(\ref{eq:decoh})  assumes the neat form
of a Bloch  equation~\cite{Leo,Raffeltbook}
 \begin{equation}
 \frac{d}{d x} {\bf P} = {{\bf h}_0 \times {\bf P}} - 
 {\mathcal D} {\bf P}_T \,\ ,
 \label{eq:pol}
 \end{equation}
 where ${\bf h}_0$ corresponds to 
Eq.~(\ref{eq:magn}) with $\xi=0$, and  ${\bf P}_T= {\bf P}- {\bf P}_3$ is the ``transverse''
  part of the polarization vector. In this picture,  the first term in the right hand side  
  represents the usual coherent ``precession'' of the polarization vector ${\bf P}$ around
  the vector
  ${\bf h}_0$~\cite{MS86,Leo,Kimb,Duan,Smirnov:1998cr}, 
  while the second (damping) term  
  produces a loss of coherence, i.e.,  a shortening  of the 
vector length $|{\bf P}|$~\cite{Leo,Raffeltbook}.

Using Eqs.~(\ref{eq:polar})--(\ref{eq:magn}), one can write 
Eq.~(\ref{eq:pol}) in components~\cite{Loreti:1994ry,Balan96,Benatti,Nuno96,Burg96}
\begin{eqnarray}
\frac{d}{d x}
\left( \begin{array}{c} P_1 \\ P_2 \\ P_3 \end{array} \right)
&=&  \left( \begin{array}{ccc}
 -  {\mathcal D}(x)  &     -V_0(x) + k_H c_{2 \theta}  &      0 \\
 V_0(x) - k_H c_{2 \theta}  &  -  {\mathcal D}(x)     &  - k_H s_{2 \theta} \\
 0  &  k_H s_{2 \theta}  &      0 
\end{array} \right)
\left( \begin{array}{c} P_1 \\ P_2 \\ P_3 \end{array} \right) \nonumber\\
&\equiv & {\mathcal K} 
\left( \begin{array}{c} P_1 \\ P_2 \\ P_3 \end{array} \right) \,\ ,
\label{eq:meanev}
\end{eqnarray}
 where $c_{2 \theta}\equiv \cos 2 \theta_{13}$ and 
$s_{2\theta}\equiv \sin 2 \theta_{13}$.  
The (fluctuation-averaged) survival probability for electron neutrinos 
$\nu = (1,0)^T$ is then given by
\begin{equation}
\label{Ptrace} 
P_{+}^H= \mathrm{Tr}[\rho \nu^\dagger   \nu]=\frac{1+ P_3}{2} \,\ .
\label{eq:pee3}
\end{equation}

 In general, the system of  equations [Eq.~(\ref{eq:meanev})]
  has to be solved numerically, as done
e.g. in~\cite{Loreti:1994ry,Balan96,Nuno96}. However, 
in the case of small-amplitude fluctuations, the damping term $\cal D$ in
 Eq.~(\ref{eq:meanev})
can be treated as a perturbation.  In this way, it is possible to integrate analytically
Eq.~(\ref{eq:meanev}).
Here we closely follow  the derivation presented in~\cite{Burg96}.

We start by diagonalizing the ${\mathcal K}$ matrix in Eq.~(\ref{eq:meanev}) 
at 0-th order in ${\mathcal D}$, estabilishing the ``instantaneous 
diagonal basis''
\begin{equation}
\left( \begin{array}{c} \tilde{P}_1 \\ \tilde{P}_2 \\ \tilde{P}_3 \end{array} \right)
= {\mathcal U} 
\left( \begin{array}{c} P_1 \\ P_2 \\ P_3 \end{array} \right) \,\ .
\end{equation}
The matrix ${\cal U}=[{\bf\hat u}_{1},
{\bf\hat u}_{2},{\bf\hat u}_0]$ diagonalizes  ${\mathcal K}$ 
as
\begin{equation}
{\mathcal K_d} = \textrm{diag}(\lambda_{1}, \lambda_{2}, \lambda_{0}) =
\mathcal{U} \mathcal{K} \mathcal{U}^\dagger \,\ ,
\end{equation}
 where the three eigenvalues are
\begin{eqnarray}
\lambda_0 &=& 0 \,\ , \label{la0} \\
\lambda_{1,2} &=& \pm i|{\bf h}_0| = 
\pm i \sqrt{(k_H\cos2\theta_{13}-V)^2+(k_H\sin2\theta_{13})^2}\,\ , \label{lp}
\end{eqnarray}
and  the corresponding eigenvectors are
\begin{eqnarray}
{\bf\hat u}_0&=&{\bf\hat h}_0 \,\ , \label{u0}\\
{\bf\hat u}_{1,2}&=&\frac{1}{\sqrt{2}}\left({\bf\hat h}_0
\times{\bf\hat s}\pm i {\bf\hat s}\right) \,\ , \label{upm}
\end{eqnarray}
with ${\bf\hat h}_0={\bf h}_0/|{\bf h}_0|$, while ${\bf\hat s}$ 
is an arbitrary unitary vector perpendicular to ${\bf h}_0$. 
By choosing  ${\bf\hat s}=(0,1,0)$,
one obtains 
\begin{equation}
{\bf\hat u}_0 = 
\left( 
\begin{array}{c} 
\sin 2 {\tilde \theta}_{13}^+  \\
0 \\
- \cos   2 {\tilde \theta}_{13}^+  
\end{array}
\right ) ,
{\bf\hat u}_1 = \frac{1}{\sqrt{2}}
\left( 
\begin{array}{c} 
\cos 2 {\tilde \theta}_{13}^+  \\
i \\
 \sin   2 {\tilde \theta}_{13}^+  
\end{array}
\right ) ,
{\bf\hat u}_2 = \frac{1}{\sqrt{2}}
\left( 
\begin{array}{c} 
\cos 2 {\tilde \theta}_{13}^+  \\
-i \\
 \sin   2 {\tilde \theta}_{13}^+  
\end{array}
\right ) 
\end{equation}
while the matrix 
 ${\cal U}$ takes the form
\begin{equation}
{\cal U}=\left[\begin{array}{ccc}
\frac{1}{\sqrt{2}}\cos2\tilde\theta_{13}^{+} & \frac{1}{\sqrt{2}}\cos2\tilde\theta_{13}^{+} &
\sin2\tilde\theta_{13}^{+} \\
\frac{i}{\sqrt{2}} & -\frac{i}{\sqrt{2}} & 0\\
\frac{1}{\sqrt{2}}\sin2\tilde\theta_{13}^{+} & \frac{1}{\sqrt{2}}\sin2\tilde\theta_{13}^{+} &
-\cos2\tilde\theta_{13}^{+}
\end{array} \right]\,\ ,
\end{equation}
where $\tilde\theta_{13}^+$ is the mixing angle in matter defined by
\begin{eqnarray}
  \sin 2{\tilde\theta}_{13}^{+} &=& \frac{\sin2\theta_{13}}
  {\sqrt{(\cos2\theta_{13} -V/k_H)^2+(\sin2\theta_{13})^2}}\ , \label{sinthetama} \\
  \cos 2{\tilde\theta}_{13}^+ &=& \frac{\cos2\theta_{13} -V/k_H}
  {\sqrt{(\cos2\theta_{13} -V/k_H )^2+(\sin2\theta_{13})^2}}\ . \label{costhetama} 
\end{eqnarray}

In the diagonal basis, Eq.~(\ref{eq:meanev}) becomes
\begin{equation}
\frac{d}{dx} \tilde{P} = {\mathcal K_d} {\tilde P} - {\mathcal U}
\frac{d\ {\mathcal U}^\dagger}{dx} {\tilde P} \,\ .
\label{eq:diag}
\end{equation}
We assume that the propagation is adiabatic ($d{\mathcal U}^{\dagger}/dx =0$), 
except near one ``crossing point''  $x_c$ where $V(x_c)\simeq k_H$ (The generalization 
to multiple crossings is straightforward). 
In this approximation, Eq.~(\ref{eq:diag}) has thus a formal solution:
\begin{equation}
{\bf P}(x)={\cal U}^\dagger(x) \,
{\cal S}(x_c,x) \, {\cal Q}(x_c) \, {\cal S}(x_0,x_c) \,
{\cal U} (x_0) \, {\bf P}(x_0)\,\ ,
\label{eq:formal}\end{equation}
where $x_0$ is the neutrino production point, and
\begin{equation}
{\cal S}(x_1,x_2)=\mathrm{diag}\left[e^{i\phi(x_1,x_2)}, e^{-i\phi(x_1,x_2)},1\right]\,\ ,
\end{equation}
is the adiabatic evolution operator, while $\phi(x_1,x_2)=\int_{x_1}^{x_2}\, dy|{\bf h}_0(y)|$, 
and
the non-diagonal operator 
${\cal Q}(x_c)$ takes into account the nonadiabatic transition (level crossing)
 around the point $x_c$. 

By inserting the above expressions in Eq.~(\ref{Ptrace}), and averaging  over the oscillatory terms, 
one gets:
\begin{equation}
P^H_{+}=\frac{1+[{\cal Q}(x_c)]_{33}\cos2\tilde\theta_{13}^+(x)\cos2\tilde\theta_{13}^+(x_0)}{2}\,\ ,
\label{eq:phpq}
\end{equation}
The comparison of Eq.~(\ref{eq:phpq}) with Eq.~(\ref{parke}) 
leads to the identification   $ [{\cal Q}(x_c)]_{33}=1 - 2 P_c$,
 where $P_c$ is the crossing probability between the two mass eigenstates in matter.
At 0-th order in ${\mathcal D}$ one thus recovers the usual (no-fluctuation) Parke's formula:
\begin{equation}
P^H_{+}=\frac{1}{2}+\left(\frac{1}{2}-P_c\right)\cos2\tilde\theta_{13}^+(x)
\cos2\tilde\theta_{13}^+(x_0)\,\ .
\end{equation}

The first-order corrections to the eigenvalues [Eqs.~(\ref{la0})--(\ref{lp})] 
and eigenvectors [Eqs.~(\ref{u0})--(\ref{upm})] are
 calculated through the standard perturbation theory as:
\begin{eqnarray}
\lambda^{(\prime)}_a&=& \lambda_a + \delta \lambda_a = \lambda_a - {\cal D}
{\bf\hat u}_{a}^\dagger {\cal G} {\bf\hat u}_{a}   \ ,\\
{\bf\hat u}_a^{(\prime)}&=& {\bf\hat u}_a + \delta {\bf\hat u}_a   = 
{\bf\hat u}_a  - {\cal D}
\sum_{b\neq a}\frac{{\bf\hat u}^\dagger _{b} {\cal G}{\bf\hat u}_{a}}{\lambda_a-\lambda_b}{\bf\hat u} _{b} \ ,
\label{eigenc}
\end{eqnarray}
where $a,b=0, 1, 2$ and ${\cal G} = \textrm{diag}(1,1,0)$. Since 
${\bf\hat u}^\dagger _{b} {\cal G}{\bf\hat u}_{a} = \delta_{ab}- ({\bf\hat u}_{a})_3({\bf\hat u}_{b})_3$, one obtains
\begin{eqnarray}
\lambda_0^\prime &=&-{\cal D}\sin^2 2\tilde\theta_{13}^+ \,\nonumber \\
\lambda_{1,2}^\prime &=& \pm i  |{\bf h}_0|
-{\mathcal D}\left(1 - \frac{1}{2} \sin^2 2 \tilde{\theta}_{13}^+\right) \,\ , 
\end{eqnarray}
and
\begin{eqnarray}
{\bf\hat u}_0^{(\prime)}&=& {\bf\hat u}_0
+ {\mathcal D}\frac{\sin 2 \tilde{\theta}_{13}^+ \cos 2 {\tilde \theta}_{13}^+}{|{\bf h}_0|}
 {\bf\hat s} \,\ , \nonumber \\
{\bf\hat u}_{1,2}^{(\prime)}&=& {\bf\hat u}_{1,2}
\mp i{\mathcal D}\frac{\sin 2 \tilde{\theta}_{13}^+}{\sqrt{2}|{\bf h}_0|}
\left[-\cos 2 \tilde{\theta}_{13}^+ {\bf\hat u}_0 +
\frac{ \sin 2\tilde{\theta}_{13}^+}{2\sqrt{2}}{\bf\hat u}_{2,1}
\right] \,\ .
\label{eigcorr}
\end{eqnarray} 
By substituiting the first-order eigenvalues in Eq.~(\ref{eq:formal}),
 the matrix ${\cal S}(x_1,x_2)$ can be rewritten as:
\begin{equation}
{\cal S}(x_1,x_2)=\mathrm{diag}\left[
e^{i\phi(x_1,x_2) + \int_{x_1}^{x_2}dx\, \delta \lambda_1}, \,
e^{-i\phi(x_1,x_2) + \int_{x_1}^{x_2}dx\, \delta \lambda_2}, \,
e^{+\int_{x_1}^{x_2}dx\, \delta\lambda_0}
\right]\,\ .
\end{equation}
The expressions of $\delta\lambda_{1,2}$ are irrelevant for our pourposes, since
they disappear after averaging over the oscillating terms in $P^H_+$.
Notice that, in Eq.~(\ref{eq:formal}), only the operator  ${\cal S}$ 
is corrected at first order. The matrix ${\mathcal U}(x_0)$ remains 
unchanged, since the high-density condition 
 $V(x_0)/k_H \gg 1$ implies that $\sin 2 \tilde{\theta}_{13}^+(x_0)\simeq 0$
in Eq.~(\ref{eigcorr}).
The matrix $\mathcal{U}(x)$ at the detection point is also unchanged, since
 ${\tilde\theta}_{13}^+(x)= \theta_{13}$ and ${\mathcal D}=0$ at the Earth.
 
Finally, for small fluctuations, we assume that the operator 
 ${\cal Q}(x_c)$ (which only depends on the crossing condition
$V(x_c)\simeq k_H$) is unchanged. 
After averaging on the oscillating terms one then obtains
\begin{equation}
P^H_{+}\simeq \frac{1}{2}+\left(\frac{1}{2}-P_c\right)e^{-\Gamma_+}\cos2\tilde\theta_{13}^+(x)
\cos2\tilde\theta_{13}^+(x_0)\,\ ,
\label{eq:nonadiab}\end{equation}
where 
\begin{equation}
\Gamma_+=\int_{x_0}^{x} dy\lambda_0^\prime(y)=\int_ {x_0}^{x}dy {\cal D}(y)
\sin^2 2\tilde\theta_{13}^+(y) \,\ .
\label{eq:gamma}\end{equation}

In the case of detected supernova neutrinos, ${\tilde\theta}_{13}^+(x)$ corresponds to the vacuum value
$\theta_{13}$, while the initial high-density condition 
implies $\cos 2 {\tilde \theta}_{13}^+(x_0) \simeq -1$. With these positions,
 Eq.~(\ref{eq:nonadiab}) gives the desired expression
\begin{equation}  
P^H_+ \simeq  \frac{1}{2}- \left( \frac{1}{2}-P_c \right) 
e^{-\Gamma_+} \cos 2 \theta_{13} \,\ .
\label{nadiabappr}
\end{equation}

\begin{figure}[t]
\vspace*{-0.0cm}\hspace*{-0.2cm}
\begin{center}
\includegraphics[scale=0.6]{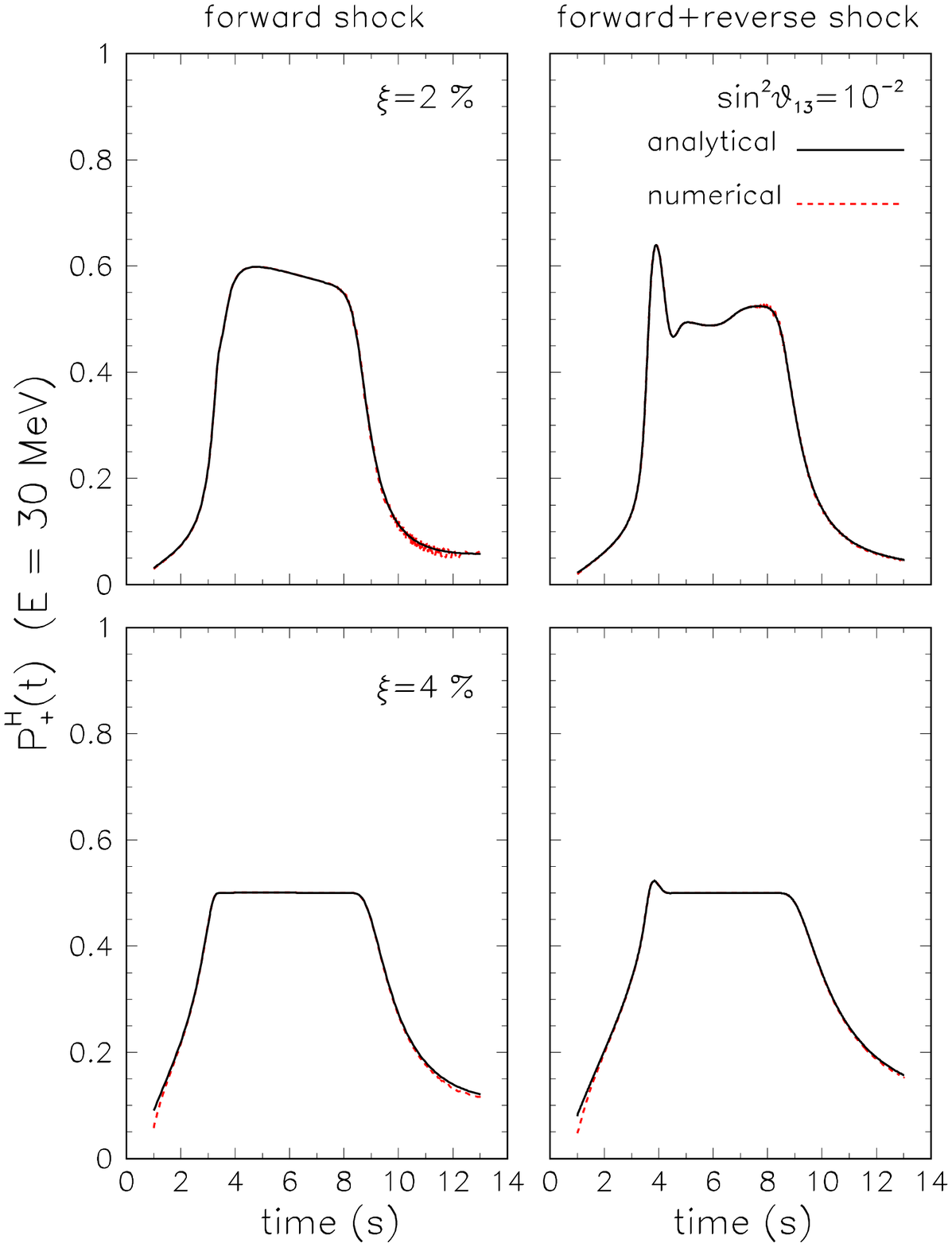}
\vspace*{+0.2cm} \caption{\label{num} Comparison of analytical and
numerical calculations of the electron neutrino survival probability
$P_{+}^H$ in the $H$ subsystem (solid and dashed curves, respectively) for $E=30$~MeV,  
$\sin^2{\theta_{13}}= 10^{-2}$, and fluctuations with  amplitude
$\xi=2 \%$ (upper panels) and $\xi=4 \%$ (lower panels).
The left panels refer to the case of forward shock only, while the
right panels to the case of forward plus reverse shock.
\footnotesize\baselineskip=4mm }
\end{center}
\end{figure}

 In order to check the reliability of  the analytical approximation
of  $P_{+}^H$ in  Eq.~(\ref{nadiabappr}), we have  compared  
it with the results of a numerical (Runge-Kutta) evolution of neutrino
master equation [Eq.~(\ref{eq:meanev})] along representative fluctuating shock-wave
profiles.  
Figure~\ref{num} shows our calculation of $P_{+}^H(t)$ at fixed neutrino energy
($E=30$~MeV) for $\sin^2 \theta_{13}=10^{-2}$, and for two representative 
fluctuation amplitudes,  $\xi=2 \%$ (upper panels) and $\xi=4\%$ (lower panels). We consider
both the case of  forward shock only (left panels) and forward plus reverse shock (right panels).  
It can be seen that the analytical calculations (solid curves)
and the numerical Runge-Kutta calculations (dashed curves) are in very good agreement
in the whole time interval 
$t \in [1,13]$~s.
The results of Figure~\ref{num} (and of other checks that we have performed
for different values of $\sin^2 \theta_{13}$ and of the neutrino energy) show that the analytical 
calculation of  $P_{+}^H$  reproduces the numerical results with high accuracy.
The same reassuring check has been performed also in the case of $P^H_-$ (not shown).

\clearpage

\section*{References} 


\begin{thebibliography}{99}
\bibitem{review}
  K.~Kotake, K.~Sato and K.~Takahashi,
  ``Explosion Mechanism, Neutrino Burst, and Gravitational Wave in
  Core-Collapse Supernovae,'' Rept.\ Prog.\ Phys.\  {\bf 69}, 971 (2006)
  [astro-ph/0509456].

  
  \bibitem{Raffrev}
  G.~G.~Raffelt,
  ``Supernova neutrino oscillations,''  
   in the Proceedings of \emph{Nobel Symposium 2004},
   Neutrino Physics (Haga Slott, Enk{\"o}ping, Sweden, 2004),
   Phys.\ Scripta {\bf T121}, 102 (2005) [hep-ph/0501049].

  
  \bibitem{Digherev}
  A.~Dighe,
  ``Supernova neutrinos: Production, propagation and oscillations,''
  in the Proceedings of \emph{Neutrino 2004},
    21st International Conference 
    on Neutrino Physics and Astrophysics (Paris, France, 14-19 Jun 2004),
     Nucl.\ Phys.\ Proc.\ Suppl.\  {\bf 143}, 449 (2005)
  [hep-ph/0409268]

\bibitem{Cavanna}
  F.~Cavanna, M.~L.~Costantini, O.~Palamara and F.~Vissani,
  ``Neutrinos as astrophysical probes,''
  Surveys High Energ.\ Phys.\  {\bf 19}, 35 (2004)
  [astro-ph/0311256].


\bibitem{Schi}
  R.~C.~Schirato and  G.~M.~Fuller,
  ``Connection between supernova shocks, flavor transformation, and the
  neutrino signal,'' astro-ph/0205390 (unpublished).

\bibitem{Taka}
  K.~Takahashi, K.~Sato, H.~E.~Dalhed and J.~R.~Wilson,
  ``Shock propagation and neutrino oscillation in supernova,''
  Astropart.\ Phys.\  {\bf 20}, 189 (2003)
  [astro-ph/0212195].

\bibitem{Luna}
  C.~Lunardini and A.~Yu.~Smirnov,
  ``Probing the neutrino mass hierarchy and the 13-mixing with supernovae,''
  JCAP {\bf 0306}, 009 (2003)
  [hep-ph/0302033].

\bibitem{Foglish}
  G.~L.~Fogli, E.~Lisi, A.~Mirizzi,  and D.~Montanino,
  ``Analysis of energy- and time-dependence of supernova shock effects on
  neutrino crossing probabilities,''
  Phys.\ Rev.\ D {\bf 68}, 033005 (2003)
  [hep-ph/0304056].
  
\bibitem{Tomas}
  R.~Tom{\`a}s, M.~Kachelrie{\ss}, G.~Raffelt, A.~Dighe, H.~T.~Janka and L.~Scheck,
  ``Neutrino signatures of supernova shock and reverse shock propagation,''
  JCAP {\bf 0409}, 015 (2004)
  [astro-ph/0407132].
  
  \bibitem{KaWa} S.~Kawagoe, ``Shock Wave Propagation 
 in Prompt Supernova Explosion and the MSW Effect of Neutrino,''
  talk at \emph{TAUP~2005}, 9th International Conference on Topics in Astroparticle
 and Underground Physics (Zaragoza, Spain, 2005), available at 
 www.unizar.es/taup2005/talks.htm.

  
  \bibitem{Dasgupta}
  B.~Dasgupta and A.~Dighe,
  ``Phase effects in neutrino conversions during a supernova shock wave,''
    hep-ph/0510219.

  
  \bibitem{FogliMega}
  G.~L.~Fogli, E.~Lisi, A.~Mirizzi and D.~Montanino,
  ``Probing supernova shock waves and neutrino flavor transitions in
  next-generation water-Cherenkov detectors,''
  JCAP {\bf 0504}, 002 (2005)
  [hep-ph/0412046].

\bibitem{Barger:2005it}
  V.~Barger, P.~Huber and D.~Marfatia,
  ``Supernova neutrinos can tell us the neutrino mass hierarchy  independently
  of flux models,''
  Phys.\ Lett.\ B {\bf 617}, 167 (2005)
  [hep-ph/0501184].


\bibitem{Choubey:2006aq}
  S.~Choubey, N.~P.~Harries and G.~G.~Ross,
  ``Probing neutrino oscillations from supernovae shock waves via the IceCube
  detector,'' hep-ph/0605255.


\bibitem{Keilconv}
  W.~Keil, H.~T.~Janka and E.~Muller,
  ``Ledoux-convection in protoneutron stars: A clue to supernova
  nucleosynthesis?,'' Astrophys.\ Journ.\ {\bf 473}, L111 (1996)
 [astro-ph/9610203].

\bibitem{Kifon}
  K.~Kifonidis, T.~Plewa, H.~T.~Janka and E.~Mueller,
  ``Non-spherical core collapse supernovae. I: Neutrino-driven convection,
  Rayleigh-Taylor instabilities, and the formation and propagation of  metal
  clumps,''
  Astron.\ Astrophys.\  {\bf 408}, 621 (2003)
  [astro-ph/0302239].

\bibitem{Bura05}
  R.~Buras, H.~T.~Janka, M.~Rampp and K.~Kifonidis,
  ``Two-Dimensional Hydrodynamic Core-Collapse Supernova Simulations with
  Spectral Neutrino Transport II. Models for Different Progenitor Stars,''
  astro-ph/0512189.
  
  \bibitem{Scheck:2006rw}
  L.~Scheck, K.~Kifonidis, H.~T.~Janka and E.~Mueller,
  ``Multidimensional Supernova Simulations with Approximative Neutrino
  Transport I. Neutron Star Kicks and the Anisotropy of Neutrino-Driven
  Explosions in Two Spatial Dimensions,'' astro-ph/0601302.


\bibitem{Sawyer}
  R.~F.~Sawyer,
  ``Neutrino oscillations in inhomogeneous matter,''
  Phys.\ Rev.\ D {\bf 42}, 3908 (1990).


\bibitem{Loreti:1994ry}
  F.~N.~Loreti and A.~B.~Balantekin,
  ``Neutrino oscillations in noisy media,''
  Phys.\ Rev.\ D {\bf 50}, 4762 (1994)
  [nucl-th/9406003].
  
  \bibitem{Balan96}
  A.~B.~Balantekin, J.~M.~Fetter and F.~N.~Loreti,
  ``The MSW effect in a fluctuating matter density,''
  Phys.\ Rev.\ D {\bf 54}, 3941 (1996)
  [astro-ph/9604061].
  
  \bibitem{Benatti}
  F.~Benatti and R.~Floreanini,
  ``Dissipative neutrino oscillations in randomly fluctuating matter,''
  Phys.\ Rev.\ D {\bf 71}, 013003 (2005)
  [hep-ph/0412311].



\bibitem{Nuno96}
  H.~Nunokawa, A.~Rossi, V.~B.~Semikoz and J.~W.~F.~Valle,
  ``The effect of random matter density perturbations on the MSW solution to
  the solar neutrino problem,''
  Nucl.\ Phys.\ B {\bf 472}, 495 (1996)
  [hep-ph/9602307].

\bibitem{Burg96}
  C.~P.~Burgess and D.~Michaud,
  ``Neutrino propagation in a fluctuating sun,''
  Annals Phys.\  {\bf 256}, 1 (1997)
  [hep-ph/9606295].
  
  \bibitem{Bykov}
  A.~A.~Bykov, M.~C.~Gonzalez-Garcia, C.~Pena-Garay, V.~Y.~Popov and V.~B.~Semikoz,
  ``MSW solutions to the solar neutrino problem in presence of noisy matter
  density fluctuations,'' hep-ph/0005244.
  
  \bibitem{Burg02}
  C.~Burgess, N.~S.~Dzhalilov, M.~Maltoni, T.~I.~Rashba, V.~B.~Semikoz, M.~Tortola and J.~W.~F.~Valle,
  ``Large mixing angle oscillations as a probe of the deep solar interior,''
  Astrophys.\ J.\  {\bf 588}, L65 (2003)
  [hep-ph/0209094].

  
  \bibitem{Guzzo}
  M.~M.~Guzzo, P.~C.~de Holanda and N.~Reggiani,
  ``Large mixing angle solution to the solar neutrino problem and random
  matter density perturbations,''
  Phys.\ Lett.\ B {\bf 569}, 45 (2003)
  [hep-ph/0303203].


\bibitem{Bal03}
  A.~B.~Balantekin and H.~Yuksel,
  ``Do the KamLAND and solar neutrino data rule out solar density
  fluctuations?,''
  Phys.\ Rev.\ D {\bf 68}, 013006 (2003)
  [hep-ph/0303169].


\bibitem{LoreSN}
  F.~N.~Loreti, Y.~Z.~Qian, G.~M.~Fuller and A.~B.~Balantekin,
  ``Effects of random density fluctuations on matter enhanced neutrino flavor
  transitions in supernovae and implications for supernova dynamics and
  nucleosynthesis,''
  Phys.\ Rev.\ D {\bf 52}, 6664 (1995)
  [astro-ph/9508106].


  
  \bibitem{Dighe}
  A.~S.~Dighe and A.~Yu.~Smirnov,
  ``Identifying the neutrino mass spectrum from the neutrino burst from a
  supernova,''
  Phys.\ Rev.\ D {\bf 62}, 033007 (2000)
  [hep-ph/9907423].

 \bibitem{LisiSN}
  G.~L.~Fogli, E.~Lisi, D.~Montanino and A.~Palazzo,
  ``Supernova neutrino oscillations: A simple analytical approach,''
  Phys.\ Rev.\ D {\bf 65}, 073008 (2002)
  [Erratum-ibid.\ D {\bf 66}, 039901 (2002)]
  [hep-ph/0111199].
  
  
  \bibitem{Fogl2005}
  G.~L.~Fogli, E.~Lisi, A.~Marrone and A.~Palazzo,
  ``Global analysis of three-flavor neutrino masses and mixings,'' hep-ph/0506083;
  to appear in Prog.\ Part.\ Nucl.\ Phys (2006).

  
 \bibitem{CHOOZ}
  M.~Apollonio {\it et al.},
  ``Search for neutrino oscillations on a long base-line at the CHOOZ  nuclear
  power station,''
  Eur.\ Phys.\ J.\ C {\bf 27}, 331 (2003)
  [hep-ex/0301017].
  
   \bibitem{KuoSN}
  T.~K.~Kuo and J.~T.~Pantaleone,
  ``Supernova Neutrinos And Their Oscillations,''
  Phys.\ Rev.\ D {\bf 37}, 298 (1988).

 \bibitem{MikSmir}
  S.~P.~Mikheyev and A.~Yu.~Smirnov,
  ``Resonant Neutrino Oscillations In Matter,''
  Prog.\ Part.\ Nucl.\ Phys.\  {\bf 23}, 41 (1989).
  
  \bibitem{KuPa}
  T.~K.~Kuo and J.~T.~Pantaleone,
  ``Neutrino Oscillations In Matter,''
  Rev.\ Mod.\ Phys.\  {\bf 61}, 937 (1989).

  
  
 

 \bibitem{Matt}  L.~Wolfenstein,
				``Neutrino Oscillations In Matter,''
                Phys.\ Rev.\ D {\bf 17}, 2369 (1978);
                S. P.~Mikheev and A. Yu.\ Smirnov,
                ``Resonance Enhancement Of Oscillations In Matter And Solar Neutrino
				Spectroscopy,''
                Yad.\ Fiz.\ {\bf 42}, 1441 (1985)
                [Sov.\ J.\ Nucl.\ Phys.\ {\bf 42}, 913 (1985)].
                
    \bibitem{Petcov}
  S.~T.~Petcov,
  ``Exact Analytic Description Of Two Neutrino Oscillations In Matter With
  Exponentially Varying Density,''
  Phys.\ Lett.\ B {\bf 200}, 373 (1988).
  
 \bibitem{landau}
L.~D.~Landau,
``Zur Theorie der Energie{\"u}bertragung bei St{\"o}ssen''
{\em(``On the theory of the energy transfer in collisions'')\/},
 Phys.\ Z.\ Sowjetunion {\bf 1}, 88 (1932); 
 ``Zur Theorie der Energie{\"u}bertragung, II'' {\em(``On the theory of the energy
transfer, II'')\/},
    Phys.\ Z.\ Sowjetunion {\bf 2}, 46 (1932).

 \bibitem{zener}
  C.~Zener,
  ``Nonadiabatic Crossing Of Energy Levels,''
  Proc.\ Roy.\ Soc.\ Lond.\ A {\bf 137}, 696 (1932).
  
  \bibitem{stuck}
  E.~C.~St{\"u}ckelberg,
``Theorie der unelastichen St{\"o}sse zwischen Atomen''
{\em(``Theory of the inelastic collisions between atoms'')\/},
 Helv.\ Phys.\ Acta {\bf 5}, 369 (1932).
  
  \bibitem{majo}
  E.~Majorana, ``Atomi orientati in campo magnetico
 variabile'' {\em(``Oriented atoms 
in variable magnetic field'')\/}, Nuovo Cimento {\bf 9}, 43 (1932).
  
  \bibitem{digiac}
  
  F.~di Giacomo and E.~E.~Nikitin,
 `` Majorana formula and the Landau-Zener-St{\"u}ckelberg treatment of the avoided
   crossing problem,''
 Ups.\ Fiz.\ Nauk {\bf 175} (5), 545 (2005)
  [Physics-Uspekhi {\bf 48} (5), 515 (2005)].
  

  \bibitem{Parke}
  S.~J.~Parke,
  ``Nonadiabatic Level Crossing in Resonant Neutrino Oscillations,''
  Phys.\ Rev.\ Lett.\  {\bf 57}, 1275 (1986). 
 
\bibitem{SKde}	
   Y.~Fukuda {\it et al.},
  ``The Super-Kamiokande detector,''
  Nucl.\ Instrum.\ Meth.\ A {\bf 501}, 418 (2003).



\bibitem{UNNO}	UNO official homepage: ale.physics.sunysb.edu	

\bibitem{Jung}	C.~K.~Jung,
  ``Feasibility of a next generation underground water Cherenkov detector:
  UNO,'' hep-ex/0005046.
	

\bibitem{HK03}	K.~Nakamura,
  ``Hyper-Kamiokande: A next generation water Cherenkov detector,''
  Int.\ J.\ Mod.\ Phys.\ A {\bf 18}, 4053 (2003).

\bibitem{Vill}	L.\ Mosca, 
``Status of the project of a Large International Underground Laboratory
at Frejus,''
talk at the Villars CERN/SPSC Meeting	
				(Villars, Switzerland, 2004), available at
				nuspp.in2p3.fr/Frejus. See also 
				 J.~E.~Campagne, M.~Maltoni, M.~Mezzetto and T.~Schwetz,
  ``Physics potential of the CERN-MEMPHYS neutrino oscillation project,''
    hep-ph/0603172.

				
\bibitem{Cadonati}
  L.~Cadonati, F.~P.~Calaprice and M.~C.~Chen,
  ``Supernova neutrino detection in Borexino,''
  Astropart.\ Phys.\  {\bf 16}, 361 (2002)
  [hep-ph/0012082].
  
  \bibitem{Aglietta}
  M.~Aglietta {\it et al.},
  ``The most powerful scintillator supernovae detector: LVD,''
  Nuovo Cim.\ A {\bf 105}, 1793 (1992).


\bibitem{ober}		
 L.~Oberauer,
  ``Low energy neutrino physics after SNO and KamLAND,''
  Mod.\ Phys.\ Lett.\ A {\bf 19}, 337 (2004)
  [hep-ph/0402162]; L.~Oberauer, ``LENA,'' talk at \emph{NO-VE 2006},
  3rd International Workshop on Neutrino Oscillations in Venice (Venice, Italy, 2006),
  available at http://axpd24.pd.infn.it /NO-VE2006 /talks/NOVE$\_$Oberauer.ppt.
  
  \bibitem{Gil-Botella:2004bv}
  I.~Gil-Botella and A.~Rubbia,
  ``Decoupling supernova and neutrino oscillation physics with LAr TPC
  detectors,''
  JCAP {\bf 0408}, 001 (2004)
  [hep-ph/0404151].
  
  
  \bibitem{Mirizzi:2005tg}
  A.~Mirizzi and G.~G.~Raffelt,
  ``New analysis of the SN 1987A neutrinos with a flexible spectral shape,''
  Phys.\ Rev.\ D {\bf 72}, 063001 (2005)
  [astro-ph/0508612].

 \bibitem{Malek:2002ns}
  M.~Malek {\it et al.}  [Super-Kamiokande Collaboration],
  ``Search for supernova relic neutrinos at Super-Kamiokande,''
  Phys.\ Rev.\ Lett.\  {\bf 90}, 061101 (2003)
  [hep-ex/0209028].
  
  \bibitem{Foglirelic}
  G.~L.~Fogli, E.~Lisi, A.~Mirizzi and D.~Montanino,
  ``Three-generation flavor transitions and decays of supernova relic
  neutrinos,''
  Phys.\ Rev.\ D {\bf 70}, 013001 (2004)
  [hep-ph/0401227].
  
  \bibitem{Yuksel:2005ae}
  H.~Yuksel, S.~Ando and J.~F.~Beacom,
  ``Direct measurement of supernova neutrino emission parameters with a
  gadolinium enhanced Super-Kamiokande detector,'' astro-ph/0509297.

\bibitem{Minakata:2000rx}
  H.~Minakata and H.~Nunokawa,
  ``Inverted hierarchy of neutrino masses disfavored by supernova 1987A,''
  Phys.\ Lett.\ B {\bf 504}, 301 (2001)
  [hep-ph/0010240].


 \bibitem{MS86}
S. P. Mikheyev and A. Yu. Smirnov, ``Neutrino Oscillations In Matter With Varying Density,''
  in the Proceedings of {\it '86 Massive Neutrinos in
Astrophysics and in Particle Physics},  Sixth
Moriond Workshop (Tignes, France, 1986), edited by O. Fackler and J. Tran Thanh
Van (Editions Fronti\`eres, Gif-sur-Yvette, 1986), p. 355.

\bibitem{Leo}
  L.~Stodolsky,
  ``On The Treatment Of Neutrino Oscillations In A Thermal Environment,''
  Phys.\ Rev.\ D {\bf 36}, 2273 (1987).
  
  
 \bibitem{Kimb}
  C.~W.~Kim, J.~Kim and W.~K.~Sze,
  ``On The Geometrical Representation Of Neutrino Oscillations In Vacuum And
  Matter,''
  Phys.\ Rev.\ D {\bf 37}, 1072 (1988).
  
  \bibitem{Duan}
  H.~Duan, G.~M.~Fuller and Y.~Z.~Qian,
  ``Collective neutrino flavor transformation in supernovae,'' astro-ph/0511275.

\bibitem{Lindblad:1975ef}
  G.~Lindblad,
  ``On The Generators Of Quantum Dynamical Semigroups,''
  Commun.\ Math.\ Phys.\  {\bf 48}, 119 (1976).

\bibitem{Smirnov:1998cr}
  A.~Yu.~Smirnov,
  ``Neutrino conversion and neutrino astrophysics,''
  hep-ph/9811296, in the Proceedings of 
 \emph{Symposium on New Era in Neutrino Physics} (Tokyo, Japan, 1998),
published in  ``Tokyo 1998, New era in neutrino physics'', 
edited by H. Minakata and O. Yasuda,
 Frontiers of Science Series n. 25 (Univ. Academy Press, Tokyo,  
Japan, 1999), p 1.

\bibitem{Raffeltbook}
  G.~G.~Raffelt,
   ``Stars as Laboratories for Fundamental Physics,''
  (Univ.\ of Chicago Press, Chicago, IL, 1996).
  
 





\end{thebibliography}
\end{document}